\documentclass[amsmath,amssymb]{revtex4}

\usepackage{graphicx}
\usepackage{dcolumn}
\usepackage{bm}

\usepackage{amsfonts}
\usepackage{graphicx, color}
\usepackage{hyperref}

\begin{document}


\title[QAC and TSP]{Quantum Adiabatic Computation and the Travelling Salesman Problem}


\author{Tien D. Kieu}
\email{kieu@swin.edu.au}
\affiliation{Centre for Atom Optics and Ultrafast Spectroscopy, ARC
Centre of Excellence for Quantum-Atom Optics, Swinburne University
of Technology, Hawthorn 3122, Australia}




\begin{abstract}
The NP-complete problem of the travelling salesman (TSP) is
considered in the framework of quantum adiabatic computation (QAC).
We first derive a remarkable lower bound for the computation time
for adiabatic algorithms in general as a function of the energy
involved in the computation.	Energy, and not just time and space,
must thus be considered in the evaluation of algorithm complexity,
in perfect accordance with the understanding that all computation is
physical. We then propose, with oracular Hamiltonians, new quantum
adiabatic algorithms of which not only the lower bound in time but
also the energy requirement do not increase exponentially in the
size of the input. Such an improvement in both time and energy
complexity, as compared to all other existing algorithms for TSP, is
apparently due to quantum entanglement. We also appeal to the
general theory of Diophantine equations in a speculation on physical
implementation of those oracular Hamiltonians.
\end{abstract}

\maketitle

\section*{A lower bound on the adiabatic computation time}
Right from the earlier days of quantum computation, NP-complete
problems have been the subject of investigation within standard
quantum mechanics~\cite{Cerny1993} as well as in non-standard
theory~\cite{Abrams1998}.  Then, with the advent of QAC~\cite{qac}
they are further pursued in the form of satisfiability
problems~\cite{Farhib, Farhia, Farhi2002} mostly with inconclusive
results~\cite{Znidaric2005, Mitchell2005} (see also~\cite{Farhi2005,
Znidaric2005a}).	QAC has also been applied to the search of an
unstructured database with the same time complexity as with
Grover's~\cite{Grover, Roland2002}, or with a much shorter time but
at a more substantial cost for the energy
involved~\cite{Eryigit2003, Das2003, Wei}.

QAC starts with the readily constructible ground state $|g_I\rangle$
of an initial Hamiltonian $H_I$ which is then {\em adiabatically}
extrapolated to the final Hamiltonian $H_P$ so that the ground state
$|g\rangle$ of the latter which contains the information of the
solution of the problem we want to solve could be obtained with
reasonably high probability.	The interpolation between $H_I$ and
$H_P$ is facilitated by a time-dependent Hamiltonian in the time
interval $0\le t \le T$,
\begin{eqnarray}
{\cal H}(t) &=& f(t)H_I + g(t)H_P,
\label{extrapolation}
\end{eqnarray}
either in a temporally linear manner (in which case, $f(t) =
\left(1-\frac{t}{T}\right)$ and $g(t)=\frac{t}{T} $), or otherwise
(but with $f(0) = 1 = g(T)$ and $f(T) = 0 = g(0)$), and with or
without the condition $f(t) + g(t) = 1$ for all $t$.	Such a time
evolution is captured by the Schr\"odinger equation
\begin{eqnarray}
i\partial_t |\psi(t)\rangle &=& {\cal H}(t)\;|\psi(t)\rangle,
\label{Schroedinger}\\
|\psi(0)\rangle &=& |g_I\rangle. \nonumber
\end{eqnarray}

We next consider another state $|\phi(t)\rangle$ which satisfies a
closely related Schr\"odinger equation
\begin{eqnarray}
i\partial_t |\phi(t)\rangle &=& \left(f(t)H_I+\beta g(t) {\bf 1}\right)\;|\phi(t)\rangle,
\label{Schroedingerb}\\
|\phi(0)\rangle &=& |g_I\rangle, \nonumber
\end{eqnarray}
with arbitrary $\beta$, resulting in a phase ambiguity.  Clearly up
to a phase,
\begin{eqnarray}
|\phi(t)\rangle={\rm e}^{i\xi(t)}|g_I\rangle. \label{phase}
\end{eqnarray}

From the difference between Eqs.~(\ref{Schroedinger})
and~(\ref{Schroedingerb}),
\begin{eqnarray}
\partial_t \left(|\psi(t)\rangle - |\phi(t)\rangle \right)= -i{\cal H}(t) \left(|\psi(t)\rangle -
|\phi(t)\rangle\right) + i g(t)(H_P-\beta)|\phi(t)\rangle,
\end{eqnarray}
we have
\begin{eqnarray}
\partial_t \|\,|\psi(t)\rangle - |\phi(t)\rangle \|^2 &=& - 2\Re\, \left(\langle \psi(t)| - \langle\phi(t)|\right)
\partial_t\left(|\psi(t)\rangle -
|\phi(t)\rangle\right),
\nonumber\\
&=& 2g(t)\Im\, \left(\langle \psi(t)| -
\langle\phi(t)|\right)(H_P-\beta)|\phi(t)\rangle,
\nonumber\\
&\le& 2g(t) \|(H_P-\beta)|\phi(t)\rangle\|\times\|\,|\psi(t)\rangle
- |\phi(t)\rangle\|,\nonumber\\
\partial_t \|\,|\psi(t)\rangle - |\phi(t)\rangle \| &\le& g(t)
\|(H_P-\beta)|g_I\rangle\|,
\end{eqnarray}
where the first inequality is a result of the Schwarz inequality;
and the second inequality from~(\ref{phase}). Integrating the time
variable on both sides of the last expression from $0$ to $T$ yields
\begin{eqnarray}
\frac{\|\,|\psi(T)\rangle - |\phi(T)\rangle
\|}{\|(H_P-\beta)|g_I\rangle\|} &\le& \int_0^T g(\tau) d\tau.
\label{7}
\end{eqnarray}

The numerator of the left hand side is the distance between the
final state of $|\psi(T)\rangle$ and essentially its initial state
at $t=0$,
\begin{eqnarray}
\|\,|\psi(T)\rangle - |\phi(T)\rangle \| &\le& 2.
\end{eqnarray}

Now, with the help of the following result
\begin{eqnarray}
{\|(H_P-\beta)|g_I\rangle\|} &=&
{\sqrt{\langle g_I|(H_P-\beta)^2|g_I\rangle}}, \nonumber\\
&=& {\sqrt{\langle g_I|H_P^2|g_I\rangle
-2\beta \langle g_I|H_P|g_I\rangle + \beta^2}}, \nonumber\\
&=& {\sqrt{(\langle g_I|H_P^2|g_I\rangle -\langle
g_I|H_P|g_I\rangle^2) + (\beta - \langle
g_I|H_P|g_I\rangle)^2}}, \nonumber\\
&\ge& {\sqrt{\langle g_I|H_P^2|g_I\rangle -\langle
g_I|H_P|g_I\rangle^2}} \equiv {\Delta_I E}, \label{spread}
\end{eqnarray}
we have also
\begin{eqnarray}
\frac{\||\psi(T)\rangle - |\phi(T)\rangle
\|}{\|(H_P-\beta)|g_I\rangle\|} \le
\frac{2}{\min_{\beta}\|(H_P-\beta)|g_I\rangle\|} \le
\frac{2}{\Delta_I E}.  \label{9}
\end{eqnarray}
This represents an upper bound whereby there is no restriction on
how large the distance from the final state, $|\psi(T)\rangle$, to
the initial state, $|\phi(T)\rangle = {\rm e}^{i\xi(T)}|g_I\rangle$,
must be.

Now, as the right hand side of~(\ref{7}) is a monotonically
increasing function of $T$ (thanks to the positivity of $g(\tau)$),
we can {\em choose} a minimum time $T_{\rm min}$ in such a way that
this right hand side is larger than the right hand side
of~(\ref{9}), which is independent of $T$,
\begin{eqnarray}
\frac{2}{\Delta_I E} &\le& \int_0^{T_{\rm min}} g(\tau) d\tau.
\label{last}
\end{eqnarray}
The meaning of this time $T_{\rm min}$ must then be derived from the
meaning of the right hand side of~(\ref{9}).  Since this latter
quantity represents an {\em unhindered} and {\em maximally
allowable} exploration by the state vector $|\psi(T)\rangle$ of the
{\em whole} Hilbert space, we thus see that $T_{\rm min}$ is
accordingly the minimum time for such an unhindered exploration, in
order to realise in general the full potential of the adiabatic
computation. That is, $T_{\rm min}$ is a lower bound on the
evolution time whereby the whole Hilbert space can be explored
unhindered, so that the final ground state, in particular, may be
obtained with some high probability.

Of course, the computation could be run for a time $T$ less than
this $T_{\rm min}$, but then sufficient computation time has not in
general been given for a {\em full} exploration of the underlying
Hilbert space. For any run time less than $T_{\rm min}$, the state
$|\psi(T)\rangle$ can explore only some {\em smaller} portion of the
Hilbert space. (As can be seen from~(\ref{9}), this portion is
defined by a restrictive constraint on the distance from the state
at time $T$, $|\psi(T)\rangle$, to its initial state, $|g_I \rangle
= {\rm e}^{-i\xi(T)}|\phi(T)\rangle$.)

The last inequality~(\ref{last}) could also be expressed in the form
of a time-energy relation,
\begin{eqnarray}
2 \le g(\theta)\; T_{\rm min} \; \Delta_I E, \label{inequality}
\end{eqnarray}
with some $\theta$ in the range $0<\theta<T_{\rm min}$, thanks to
the mean value theorem.	 This is our first main
result~\footnote{Similar results can also be derived for a
generalised QAC~\cite{Farhi2002, Wei, Boulatov2005}, where an extra
term of the form $h(t) H_E$ (with $h(0)=h(T)=0$) is added
to~(\ref{extrapolation}) to represent some further freedom in the
adiabatic paths.}.

This lower bound on the computation time incorporates the initial
ground state $|g_I\rangle$ and the spectrum of the final Hamiltonian
together in $\Delta_I E$, which is defined in~(\ref{spread}) as the
energy spread of the initial state in terms of the final energy. The
manner of the time extrapolation is further reflected in
$g(\theta)$.  The condition~(\ref{inequality}) states that the more
the spread of the initial state in energy with respect to the final
Hamiltonian, the less the lower bound on the running time.

It thus also emphasises the fact that energy must be considered in
the running of QAC, and perhaps in all physical computation if they
are indeed all physical in the end. It is interesting to further
note that recent results obtained in~\cite{Farhi2005} reflect well
in the condition above.

This condition~(\ref{inequality}) must also contain in it the
information about the energy gap between the instantaneous ground
state and the first excited state, which in turn determines a lower
bound on the computation time as indicated by the quantum adiabatic
theorem. The dependence on the gap is not manifest but once again
hidden in $\Delta_I E$, the energy spread of the initial ground
state in terms of the energy eigenstates of the final Hamiltonian.
This is not so surprising because~(\ref{inequality}) is derived
directly from the Schr\"odinger equation, which is also the starting
point for the derivation of any version of the quantum adiabatic
theorem. We will see in the quantum adiabatic algorithms of the next
Section that such an implicit gap dependence is born out in the
perfect agreement of the lower bound~(\ref{inequality}) with all
other lower bounds obtained directly from the energy gaps, if and
when such gaps can be obtained.

\section*{Energy in the consideration of algorithm complexity}
We now apply our lower bound to various quantum adiabatic algorithms
in the literature.

In order to find the item labeled by $|m\rangle$ in an unstructured
database with $N$ items, one method~\cite{Roland2002} is to employ
the initial Hamiltonian $H_I = 1 - |g_I\rangle\langle g_I|$, with
the initial ground state $|g_I\rangle$ that has equally distritbuted
probability amplitudes among the $N$ items, $|g_I\rangle =
\frac{1}{\sqrt
 N}\sum_{i=1}^N |i\rangle$, together with the final Hamiltonian $H_P = 1 - |m\rangle\langle
 m|$, which admits $|m\rangle$ as the non-degenerate ground state.
In this case, $\Delta_I E = \sqrt{N-1}/N$.

With linear time extrapolation~\cite{Farhia} also employed in the
adiabatic computation~(\ref{extrapolation}), we substitute $g(\tau)
= \frac{\tau}{T}$ into~(\ref{inequality}) to obtain
\begin{eqnarray}
O(\sqrt N)&\le& T_{\rm min}
\end{eqnarray}
This lower bound can indeed be attained in a {\em local} adiabatic
search~\cite{Roland2002}.  Its time complexity, even though better
than a straightforward classical search, is the same as Grover's
quantum search.

Alternatively, for example, the same Hamiltonians but with different
extrapolation as proposed in~\cite{Das2003, Wei},
\begin{eqnarray}
g(\tau)=\frac{\tau}{T} +
\sqrt{N}\frac{\tau}{T}\left(1-\frac{\tau}{T}\right) \Rightarrow
\int_0^T g(\tau) d\tau \sim T\times O(\sqrt N), \label{8}
\end{eqnarray}
would lead to
\begin{eqnarray}
O(1) \le T_{\rm min},
\end{eqnarray}
that is, a {\em constant} lower bound on the computation time,
irrespective of the database size! This, however, could only be
obtained at the expense of the energy being scaled as $O(\sqrt N)$
(contained in the term $g(t)$ hidden in~(\ref{8})), even if only for
the intermediate time.

Even though being known previously~\cite{Eryigit2003, Das2003, Wei},
we want to reiterate here that energy consideration emerges
naturally from QAC, as predominantly displayed in the
relation~(\ref{inequality}), and wish to re-emphasise here the
inevitable fact that, because all computation is physical, energy
must be an independent dimension, besides those of time and space,
in the evaluation of algorithm complexity.

\section*{An adiabatic algorithm for TSP with non-polynomial energy resources}
We will deal with the version of TSP~\cite{Lewis1981} with $M$
cities to be visited as a search problem for the shortest tour among
$M!$ possible tours, each of which connects all the cities and
visits each city once only. The distances between pairs of cities
can be presented in a distance matrix, which may or may not be
symmetrical, having elements $d_{ij}$ being the (oriented) distance
from the $i$-th city to the $j$-th city, and having vanishing
diagonal elements, $d_{ii}=0$.

A tour is labeled by an integer $k$, $k = 1, \ldots, M!$, which is
the rank of a permutation of $\langle 1, \ldots, M\rangle$. The
length of a tour is labeled by $l_k$, which is the sum of
appropriate $d_{ij}$.  We also denote  by $l_{\rm max}$ a constant
(not much) greater than the maximum tour length.

We state here an assumption (needed for this Section only) about the
variance $\Sigma_M$ in $l_k$,
\begin{eqnarray}
\Sigma_M \equiv \sqrt{\frac{1}{M!}\sum_{k=1}^{M!}\left(l_k - \bar
l\right)^2} \;\; {\mbox{\rm  decreases, if at all, less than
exponentially in $M$,}} \label{variance}
\end{eqnarray}
for asymptotically large $M$. That is, we expect that when we add
one more city to the list to be visited, even though the number of
possibilities now shoots up to $(M+1)!$, the difference/spread
between the longest and the shortest tour lengths should not
decrease exponentially.	 Indeed, with randomly distributed $d_{ij}$
we can, thanks to the central limit theorem, show that $\Sigma_M\sim
O(\sqrt M)$ for $M\to\infty$, namely, as $\Sigma_M$ does increase,
the assumption~(\ref{variance}) is satisfied. We suspect that the
assumption can be proven in the Euclidean version of the TSP, which
is also an NP-complete problem~\cite{Papadimitriou1977}. In fact, it
would be sufficient for our purpose if the TSP with the
condition~(\ref{variance}) is provably NP-complete still.

To employ QAC, we introduce an oracular Hamiltonian $H_O$, of
infinite dimensions if necessary, which is diagonalised in the
number states $|n\rangle$,
\begin{eqnarray}
\langle n|H_O|n \rangle &=& \left\{
\begin{array}{ll}
l_{n+1}, &
{\mbox{\rm for $0\le n\le M!-1$,}}\\
l_{\rm max}, & {\mbox {\rm for $M! \le n$}}.
\end{array}
\right.
\label{oracular}
\end{eqnarray}
Possible physical implementation of this oracular/hypothetical
Hamiltonian will be discussed later.

We now need to find the ground state $|n_0\rangle$ of $H_O$ in order
to obtain the tour information from $n_0$ and the least tour length
from the ground state energy, assuming no degeneracy.	 Following a
quantum adiabatic algorithm introduced for Hilbert's tenth
problem~\cite{kieu-contphys, kieu-intjtheo, kieu-royal, kieu-spie,
kieuFull}, we employ a linear time interpolation from the following
initial Hamiltonian, with annihilation (creation) operators $a$
($a^\dagger$),
\begin{eqnarray}
H_I &=& (a^\dagger - \alpha^*)(a - \alpha),
\label{H1}
\end{eqnarray}
which admits as the ground state the coherent state $ |\alpha\rangle
= {\rm e}^{-\frac{|\alpha|^2}{2}}\sum_{n=0}^\infty
\frac{\alpha^n}{\sqrt{n!}}\,|n\rangle.$ The choice of the complex
number $\alpha$ will be crucial for the complexity of our algorithm.

The initial ground state $|\alpha\rangle$ has a Poissonian
probability distribution among the number states $|n\rangle$, with
equal average occupation number and variance, $\bar n = \sigma^2 =
|\alpha|^2$. As the oracular Hamiltonian $H_O$~(\ref{oracular}) is
only non-constant for $n \le M!-1$, the choice
\begin{eqnarray}
|\alpha|^2 = O (M!)
\label{choice1}
\end{eqnarray}
would yield, in the initial state, a variance $\sigma^2$ in $n$ that
covers all the $M!$ possibilities, and thus would lead to $\Delta_I
E$ in the inequality~(\ref{inequality}) taking the same magnitude as
the measure	 $\Sigma _M$ of the spread in the tour lengths among the
$M!$ tours,
\begin{eqnarray}
\Delta_I E \sim O(\Sigma _M).
\end{eqnarray}
By the assumption~(\ref{variance}) and according
to~(\ref{inequality}), we now have a lower bound for the computation
time which is now of the same order as the inverse of $\Sigma_M$ and
thus cannot grow exponentially in $M$. That is, we would gain in the
computation time complexity if such a lower bound in time could be
achieved for the QAC.

However, the price we must pay for such a sub-exponential, or even
constant, growth is the exponential cost of the energy that must be
supplied, as reflected in the choice~(\ref{choice1}), $|\alpha|^2 =
O (M!)$, for the initial Hamiltonian~(\ref{H1}).  For much larger
values of $|\alpha|^2$, the variance $\sigma^2$ would be
concentrating in large values of $n$, and accordingly $\Delta_I E$
would be vanishingly small as is evident from the constant
eigenvalues of $H_O$ in~(\ref{oracular}) for $M!< n$.	 For much
smaller value of $|\alpha|^2$, the variance $\Delta_I E$ could be
concentrating in an inappropriate domain of $l_n$ which in general
cannot lead to a sub-exponential or polynomial lower bound for the
time complexity.

The situation is then similar to other algorithms of a previous
Section which gain in the time complexity	 at the expense of the
energy complexity. Fortunately, we could improve upon the situation
to reduce the energy consumption.

\section*{Another algorithm for TSP with polynomial energy resources}
Instead of encoding a tour by a single number $k$, we now employ the
$M$-tuple $\langle m_1, \ldots, m_M\rangle$.  Each $m_i$ is a
natural number and its first $M$ values (for $i$ ranges from $0$ to
$M-1$) represent the $M$ cities.  For these, there are $M^M$ tuples,
including those with $m_i=m_j$, which do not correspond to a TSP
tour. In order to put those that are not a TSP tour out of reach, we
introduce a counterpart $\tilde l_s$ of the tour length
\begin{eqnarray}
\tilde l_s = \left\{
\begin{array}{ll}
{\mbox{\rm tour length }} l_s, & {\mbox{\rm if $s$ corresponds to a
TSP tour,}}\\
d^2 + l_{\rm max}, & {\mbox{\rm otherwise,}}
\end{array}
\right. \label{counterpart}
\end{eqnarray}
where we could use, for example, the following correspondence
between $s$ and the $M$-tuples
\begin{eqnarray}
s = 1+ m_1 + m_2M +  m_3M^{2} + \ldots + m_MM^{(M-1)}. \label{map}
\end{eqnarray}
In~(\ref{counterpart}), $d$ is a random number drawn from some
arbitrarily chosen distribution that has a finite variance {\em
independent} of $M$, $ \sigma_d = C,$ so that for those $s$ that are
not a TSP tour, $\tilde l_s \ge l_{\rm max}$.  Alternatively, one
could obtain the same result by choosing $d^2=0$ for even $s$, and
$d^2= 2l_{\rm max}$ for odd $s$ (that is, $d^2 = (1 - (-1)^s)l_{\rm
max})$.

We now start our QAC with the initial coherent state
$|\psi(0)\rangle = \otimes_{i=1}^M |\alpha_i\rangle$, which is the
ground state of the initial Hamiltonian
\begin{eqnarray}
\tilde H_I &=& \sum_{i=1}^M \left(a^\dagger_i -
\alpha_i^*\right)\left(a_i - \alpha_i\right),
\label{H2}
\end{eqnarray}
which is in turn extrapolated linearly in time to an oracular
Hamiltonian that is diagonalised in the number states $|m_1 \ldots
m_M\rangle$ according to
\begin{eqnarray}
\langle m_1 \ldots m_M|\tilde H_O|m_1 \ldots m_M \rangle &=& \left\{
\begin{array}{ll}
\tilde l_{s}, &
{\mbox{\rm for $1\le s\le M^M$,}}\\
l_{\rm max}, & {\mbox {\rm for $M^M < s$}}.
\end{array}
\right. \label{oracularB}
\end{eqnarray}
The aim of our computation is once again to find the ground state of
$\tilde H_O$, with the ground state energy being the shortest tour
length.

If we choose all the complex numbers $\alpha_i$'s such that
\begin{eqnarray}
|\alpha_i|^2 = O(M),
\label{choice2}
\end{eqnarray}
then the energy requirement for~(\ref{H2}) is only of $O(M^2)$,
which is significantly lower than that of $O(M!)$ of the last
Section. Furthermore, with this choice, the variance/spread in $s$
in the initial state, as can be computed easily from~(\ref{map}),
will be of $O(M^M)$. And with this spread the energy
variance~(\ref{spread}) of the initial state in terms of the energy
of~(\ref{oracularB}) will also be of the same magnitude as the
variance in the ``tour lengths" $\tilde l_s$.  That is, for large
$M$, $\Delta_I E \to O(\sigma_d)$, which is a predetermined constant
independent of $M$, in an exponentially manner because of the
exponential decrease in the proportion of TSP tours, $(M!/M^M)
\stackrel{M\to \infty}{\longrightarrow} {\rm e}^{-M}/\sqrt{M}$.

Applying our inequality~(\ref{inequality}) here once again, we are
led a lower bound in computation time approaching a constant value
{\em independent} of the number of cities $M$. But this time the
energy cost is only quadratic in $M$. This is our second main
result.

Classical simulations of this QAC is rather expensive because of the
$M^M$ possibilities, instead of $M!$ as in the usual formulation of
TSP. Hence, the extraordinary gain above would only be possible with
quantum computation. We suspect that quantum entanglement, which has
not been fully exploited in QAC up until now, is responsible for
this remarkable improvement.

The above algorithm can also be captured in the following QAC in a
finite-dimensional Hilbert space with linear time interpolation
between
\begin{eqnarray}
H_I = 1 - |g_I\rangle\langle g_I|, &{\rm and}& H_P =
\sum_{m_1=0}^{M-1}\ldots\sum_{m_M=0}^{M-1} \tilde l_s \;|m_1\ldots
m_M\rangle\langle m_1\ldots m_M|,\\
{\rm where}\;\; |\psi(0)\rangle = |g_I\rangle &=& \bigotimes_{i=1}^M
\left(\frac{1}{\sqrt{M}}\sum_{m_i=0}^{M-1}
|m_i\rangle\right).\nonumber
\end{eqnarray}
Note that similar kind of oracular Hamiltonians is already employed
in Grover's search~\cite{Roland2002} and involves some kind of
non-local interactions, which may make their implementation more
difficult than that of quantum circuits (which only require local
interactions) but not impossible. However, the infinite-dimension
formulation above (in~(\ref{H2}) and~(\ref{oracularB})) could
exhibit explicitly the requirement of energy through the
$\alpha_i$'s in~(\ref{H2}), and may be implemented physically with
quantum optical means, among others.

\section*{Physical Hamiltonians from the oracular Hamiltonians via Diophantine equations}
For completeness, we outline here some ways forward, in principle,
to a physical implementation of the oracular/hypothetical
Hamiltonians~(\ref{oracular}, \ref{oracularB}). The oracular
Hamiltonian~(\ref{oracular}), for example, can be regarded as a
computer program, with $n$ being an input upon which $y_n = \langle
n|H_O |n\rangle$ is the output (without loss of generality, we
choose $y$ also to be natural numbers). According to the general
theory of Diophantine equations~\cite{hilbert10, Chaitin:2005},
corresponding to such a computer program (Turing machine) there is a
Diophantine equation which only has integral solution in $(x_1,
\ldots, x_K)$,
\begin{eqnarray}
D(n, y_n; x_1, \ldots, x_K) =0,
\end{eqnarray}
for some $K$, if and only if $y_n$ is indeed the output for the
input $n$.	This is a manifestation of the computational
universality of Diophantine equations.

Following~\cite{kieu-contphys, kieu-intjtheo, kieu-royal, kieu-spie,
kieuFull}, we could then try to implement the physical Hamiltonian
\begin{eqnarray}
H_P = l_{\rm max}\left(D\left(a^\dagger_n a_n, a^\dagger_y a_y;
a^\dagger_1 a_1, \ldots, a^\dagger_K a_K\right) \right)^2 +
a^\dagger_y a_y.
\label{physical}
\end{eqnarray}
Thanks to the fact that Diophantine polynomials have only integral
values, it can be shown that the ground state energy of this
Hamiltonian is the shortest tour length we are after.	 This is the
most general but certainly not the most efficient way to consider
physical implementation of the oracular Hamiltonians.	 The
appearance of extra $K$ modes in~(\ref{physical}) would require
further consideration to ensure that we still have non-exponential
complexity in both time and energy.	 These are outside the scope of
the present paper.

\section*{Summary and concluding remarks}
We derived a remarkable lower bound on the computation time for QAC
in general as a function of the energy necessary for the
computation. Energy, being a physical quantity and an inseparable
component of any physical process, emerges naturally in QAC, as
predominantly displayed in the relation~(\ref{inequality}), and is
thus an inevitable and a natural dimension in the complexity
evaluation of an algorithm, in accordance with the view that all
computation is physical.  We then proposed some new quantum
adiabatic algorithms, with oracular Hamiltonians, for the
NP-complete TSP. Both the lower bound of the computation time and
the energy required of our proposal do not scale exponentially with
the number of cities involved. Enlisting the help of general theory
of Diophantine equations, we speculated on a general way in
principle to physically implement the oracular Hamiltonians of the
algorithms. This, however, and whether such a favourable lower bound
in the computation time could be achieved would require further
detailed investigations elsewhere.

\begin{acknowledgments}
I wish to thank Hans Briegel, Wolfgang D\"ur, Daniel Gottesman,
Peter Hannaford and Toby Ord for discussions. This work has also
been supported by the Swinburne University Strategic Initiatives.
\end{acknowledgments}

\bibliography{Computability}

\begin{thebibliography}{10}

\bibitem{Cerny1993}
Vladimir Cerny.
\newblock Quantum computers and intractable ({NP}-complete) computing problems.
\newblock {\em Phys. Rev.}, A 48:116--119, 1993.

\bibitem{Abrams1998}
Daniel~S. Abrams and Seth Lloyd.
\newblock Nonlinear quantum mechanics implies polynomial-time solution for
  {NP}-complete and {P} problems.
\newblock {\em Phys.Rev.Lett.}, 81:3992--3995, 1998.

\bibitem{qac}
E.~Farhi, J.~Goldstone, S.~Gutmann, and M.~Sipser.
\newblock Quantum computation by adiabatic evolution.
\newblock {\tt ArXiv:quant-ph/0001106}, 2000.

\bibitem{Farhib}
Edward Farhi, Jeffrey Goldstone, and Sam Gutmann.
\newblock A numerical study of the performance of a quantum adiabatic evolution
  algorithm for satisfiability.
\newblock {\tt arXiv:quant-ph/0007071}, 2000.

\bibitem{Farhia}
Edward Farhi, Jeffrey Goldstone, Sam Gutmann, Joshua Lapan, Andrew Lundgren,
  and Daniel Preda.
\newblock A quantum adiabatic evolution algorithm applied to random instances
  of an {NP}-complete problem.
\newblock {\em Science}, 292:472--475, 2001.
\newblock {\tt arXiv:quant-ph/0104129}.

\bibitem{Farhi2002}
Edward Farhi, Jeffrey Goldstone, and Sam Gutmann.
\newblock Quantum adiabatic evolution algorithms with different paths.
\newblock {\tt arxiv:quant-ph/0208135}, 2002.

\bibitem{Znidaric2005}
Marko Znidaric and Martin Horvat.
\newblock Exponential complexity of an adiabatic algorithm for an {NP}-complete
  problem.
\newblock {\tt arXiv:quant-ph/0509162}, 2005.

\bibitem{Mitchell2005}
David~R. Mitchell, Christoph Adami, Waynn Lue, and Colin~P. Williams.
\newblock A random matrix model of adiabatic quantum computing.
\newblock {\em Phys. Rev. A}, 71:052324, 2005.
\newblock {\tt arXiv:quant-ph/0409088}.

\bibitem{Farhi2005}
Edward Farhi, Jeffrey Goldstone, Sam Gutmann, and Daniel Nagaj.
\newblock How to make the quantum adiabatic algorithm fail.
\newblock {\tt arXiv:quant-ph/0512159}, 2005.

\bibitem{Znidaric2005a}
Marko Znidaric.
\newblock Eigenlevel statistics of the quantum adiabatic algorithm.
\newblock {\em Phys. Rev. A}, 72:052336, 2005.
\newblock {\tt arXiv:quant-ph/0512018}.

\bibitem{Grover}
L.K. Grover.
\newblock Quantum mechanics helps in searching for a needle in a haystack.
\newblock {\em Phys. Rev. Lett.}, 79:325--328, 1997.

\bibitem{Roland2002}
Jeremie Roland and Nicolas~J. Cerf.
\newblock Quantum search by local adiabatic evolution.
\newblock {\em Phys. Rev.}, A 65:042308, 2002.

\bibitem{Eryigit2003}
Recep Eryigit, Yigit Gunduc, and Resul Eryigit.
\newblock Local adiabatic quantum search with different paths.
\newblock {\tt arXiv:quant-ph/0309201}, 2003.

\bibitem{Das2003}
Saurya Das, Randy Kobes, and Gabor Kunstatter.
\newblock Energy and efficiency of adiabatic quantum search algorithms.
\newblock {\em J. Phys. A: Math. Gen.}, 36:1--7, 2003.

\bibitem{Wei}
Zhaohui Wei and Mingsheng Ying.
\newblock Quantum search algorithm by adiabatic evolution under a priori
  probability.
\newblock {\tt arXiv:quant-ph/0412117}, 2004.

\bibitem{Lewis1981}
H.R. Lewis and C.H. Papadimitriou.
\newblock {\em Elements of the Theory of Computation}.
\newblock Prentice Hall, New Jersey, 1981.

\bibitem{Papadimitriou1977}
Christos~H. Papadimitriou.
\newblock The {E}uclidean travelling salesman problem is {NP}-complete.
\newblock {\em Theoretical Computer Science}, 4:237--244, 1977.

\bibitem{kieu-contphys}
T.D. Kieu.
\newblock Computing the non-computable.
\newblock {\em Contemporary Physics}, 44:51--77, 2003.

\bibitem{kieu-intjtheo}
T.D. Kieu.
\newblock Quantum algorithms for {H}ilbert's tenth problem.
\newblock {\em Int. J. Theor. Phys.}, 42:1451--1468, 2003.

\bibitem{kieu-royal}
T.D. Kieu.
\newblock A reformulation of {H}ilbert's tenth problem through quantum
  mechanics.
\newblock {\em Proc. Roy. Soc.}, A 460:1535--1545, 2004.

\bibitem{kieu-spie}
T.D. Kieu.
\newblock Numerical simulations of a quantum algorithm for {H}ilbert's tenth
  problem.
\newblock In Eric Donkor, Andrew~R. Pirich, and Howard~E. Brandt, editors, {\em
  Proceedings of SPIE Vol. 5105 {\it Quantum Information and Computation}},
  pages 89--95. SPIE, Bellingham, WA, 2003.

\bibitem{kieuFull}
T.D. Kieu.
\newblock Quantum adiabatic algorithm for {H}ilbert's tenth problem: I. {T}he
  algorithm.
\newblock \texttt{ArXiv:quant-ph/0310052}, 2003.

\bibitem{hilbert10}
Yuri~V. Matiyasevich.
\newblock {\em Hilbert's Tenth Problem}.
\newblock MIT Press, Cambridge, Massachussetts, 1993.

\bibitem{Chaitin:2005}
Gregory Chaitin.
\newblock {\em Meta Math! : The Quest for Omega}.
\newblock Pantheon, New York, 2005.

\bibitem{Boulatov2005}
A.~Boulatov and V.N. Smelyanskiy.
\newblock Quantum adiabatic algorithm and large spin tunnelling.
\newblock {\em Phys. Rev. A}, 71:052309, 2005.
\newblock {\tt arXiv:quant-ph/0309150}.

\end{thebibliography}
\bibliographystyle{unsrt}

\end{document}